# DETECTION AND LOCALIZATION OF CHANGE-POINTS IN HIGH-DIMENSIONAL NETWORK TRAFFIC DATA

By Céline Lévy-Leduc and François Roueff

### CNRS, LTCI and Télécom ParisTech

We propose a novel and efficient method, that we shall call *TopRank* in the following paper, for detecting change-points in high-dimensional data. This issue is of growing concern to the network security community since network anomalies such as Denial of Service (DoS) attacks lead to changes in Internet traffic. Our method consists of a data reduction stage based on record filtering, followed by a nonparametric change-point detection test based on *U*-statistics. Using this approach, we can address massive data streams and perform anomaly detection and localization on the fly. We show how it applies to some real Internet traffic provided by France-Télécom (a French Internet service provider) in the framework of the ANR-RNRT OSCAR project. This approach is very attractive since it benefits from a low computational load and is able to detect and localize several types of network anomalies. We also assess the performance of the *TopRank* algorithm using synthetic data and compare it with alternative approaches based on random aggregation.

**1. Introduction.** Recent attacks on very popular web sites such as Yahoo and eBay, leading to a disruption of services to users, have triggered an increasing interest in network anomaly detection. Typical examples include Denial of Service (DoS) attacks—a network-based attack in which agents intentionally saturate system resources—and their distributed version (DDoS). Since the aforementioned attacks represent serious threats for computer networks, developing anomaly detection systems for ensuring the defense against them has become a major concern.

Existing detection systems to deal with DoS or DDoS attacks are based on two different approaches. The first one is a signature-based approach which compares the observed patterns of the network traffic with known attack templates. If the attack belongs to the set of known attacks listed in the database, then it can be successfully detected: Snort and Bro developed by Roesch (1999) and Paxson (1999), respectively, are two examples of such anomaly detection systems. The obvious limitation of this approach is that the signature of the anomaly has to be known in advance.

The second detection system is based on statistical tools which do not require any prior information about the kind of anomalies we are faced with. As a consequence, this approach aims at detecting anomalies which do not belong to a prescribed database. It relies on the fact that anomalies in network traffic lead to abrupt changes in some network characteristics. We choose those characteristics according to the type of attacks we are looking for. These changes occur at unknown time instants and have to be detected as soon as possible. Detecting an attack in the network traffic can thus be described as a change-point detection problem, which is a classical issue in statistics. The detection can either be performed with a fixed delay (batch approach) or with a minimal average delay (sequential approach). We refer to Basseville and Nikiforov (1993), Brodsky and Darkhovsky (1993), Csörgo and Horvath (1997) and the references therein for a complete overview of the existing methods in statistical change-point detection.

The most widespread change-point detection technique in the field of network anomaly detection is the cumulated sum (CUSUM) algorithm which was first proposed by Page (1954). The CUSUM algorithm has already been used by Wang, Zhang and Shin (2002) and Siris and Papagalou (2004) for detecting DoS attacks of the TCP (Transmission Control Protocol) SYN flooding type. Such an attack consists in exploiting the TCP three-way handshake mechanism and its limitation in maintaining half-open connections. More precisely, when a server receives a SYN packet, it returns a SYN/ACK packet to the client. Until the SYN/ACK packet is acknowledged by the client, the connection remains half-opened for a period of at most the TCP connection timeout. A backlog queue builds up in the system memory of the server to maintain all half-open connections, possibly leading to a saturation of the server. In Siris and Papagalou (2004), the authors use the CUSUM algorithm to look for a change-point in the time series corresponding to the sum of received SYN packets by all the destination IP addresses which have been requested. With such an approach, it is only possible to set off an alarm when a change occurs in the aggregated series, but it is impossible to pick out the malicious flows.

Taking into account the previous definition of a TCP/SYN flooding type attack, it would be natural, in order to identify the IP addresses involved in the attack, to analyze the time series corresponding to the number of TCP/SYN packets received by each IP address, to apply to each of them a change-point detection test and to say that it is attacked if the test sets off an alarm. This idea is used in Tartakovsky et al. (2006a, 2006b), who propose a multichannel detection procedure which is a refined version of the previously quoted algorithms: it makes it possible to detect changes which occur in a channel and which could be obscured by the normal traffic in the other channels if global statistics were used. Note that these methods only focus on the number of TCP/SYN packets received by a given destination IP address and not on the number of TCP/SYN packets sent by a given source IP address to a given destination IP address. Applying a change-point detection test to the latter time series to detect an anomaly is bound to fail because of *spoofing*, a method used by attackers which makes their IP address appear as a random address on the Internet.

Operators seeking to understand and manage their networks are increasingly looking at wide-area-network traffic flows. In this framework, previous methods are bound to failure since the quantity of data to analyze is too massive. Indeed, each flow is characterized by 5 fields: source and destination IP addresses, source and destination ports and protocol number, which produces a very large database to store and study. In order to detect anomalies in such massive data streams within a reasonable time span, it is impossible to analyze the time series of all the IP addresses receiving TCP/SYN packets. That is why dimension reduction techniques have to be used. Two main approaches have been proposed: some of them are based on Principal Component Analysis (PCA) techniques [see Lakhina, Crovella and Diot (2004)] and others on random aggregation (sketches); see Krishnamurthy et al. (2003) and Li et al. (2006). Localization of the anomalies is possible with the second approach using hash table inversion techniques; see Salem, Vaton and Gravey (2007) and Abry, Borgnat and Dewaele (2007).

In this paper we propose a novel algorithm for detecting change-points in a multi-dimensional time series $(N_i^\Delta(t))_{t\geq 1}$, $i \in \{1, \ldots, D\}$, under computational constraints which allow us to process the data on the fly, even for a high dimension $D$. Our method can be used for identifying DoS and DDoS attacks as well as PortScan and NetScan in Internet traffic in cases where we are faced with massive data streams. More precisely, we can identify anomalies of the following types: TCP/SYN flooding, UDP flooding, PortScan and NetScan. UDP flooding is an attack similar to TCP/SYN flooding which aims at saturating the memory of a destination IP address by sending a lot of UDP packets. A PortScan consists in sending TCP packets to each port of a machine to know which ones are open. In a NetScan attack, a source IP address sends packets to a large number of IP addresses in order to detect the machines which are connected to the network. The problem of change-point detection in high-dimensional data mainly concerns the network security community but, in our view, it is a challenging issue which will benefit a broader audience.

Let us now give an outline of this paper. Section 2 describes the framework of our study. In Section 3 we describe the different methods that we propose to detect change-points in a multi-dimensional time series under computational constraints. These approaches are based on two different dimension reduction techniques: *TopRank* uses record filtering, whereas *HashRank* is based on random aggregation. The detection stage uses a nonparametric change-point detection test based on $U$-statistics. More precisely, we used rank tests for censored data as proposed and analyzed in Gombay and Liu (2000). The corresponding algorithms have been written in C language [for *TopRank*, see Lévy-Leduc, Benmammar and Roueff (2008)] and applied to real datasets corresponding to some Internet traffic provided by France-Télécom (a French Internet Service Provider) within the framework of the ANR-RNRT OSCAR project. In Section 4 we apply the *TopRank* algorithm to a set of real Internet traffic provided by France-Télécom which contains SYN flooding type attacks in order to explain how to choose the most relevant parameters. In Section 5 *TopRank* is applied to a set of real data containing several types of attacks: SYN flooding, UDP flooding and also PortScan and NetScan. Our research indicates that our method can be used to analyze a very large amount of data and to detect network anomalies on the fly. The methods *TopRank* and *HashRank* are finally compared first from a theoretical point of view on a toy example in Section 6 and then with Monte Carlo experiments on synthetic data in Section 7.

**2. Description of the data.** In Network applications the raw data consists of Netflow type data collected at several points of the Internet network and put into a single Netflow type table. The data at our disposal includes the source and destination IP addresses, the source and destination ports, the start time and the end time of the flow, as well as the protocol and the number of exchanged packets. In the case of the TCP protocol, the number of packets of each type (SYN, SYN/ACK, FIN, RST) is available.

Depending on the type of attack that one wants to detect, some time indexed traffic characteristics are of particular interest and have to be processed for detection purposes. For instance, in the case of the TCP/SYN flooding, the quantity of interest is the number of TCP/SYN packets received by each destination IP address per unit of time.

We denote by $\Delta$ the smallest time unit used for building time series from our raw Netflow type data. The time series are built as follows: in the case of TCP/SYN flooding, we shall denote by $N_i^\Delta(t)$ the number of TCP/SYN packets received by the destination IP address $i$ in the sub-interval indexed by $t$ of size $\Delta$ seconds. The corresponding time series of the destination IP address $i$ will thus be $(N_i^\Delta(t))_{t\geq 1}$. In the case of UDP flooding, $N_i^\Delta(t)$ will be defined as the number of UDP packets received by the destination IP address $i$ in the $t$th sub-interval of size $\Delta$ seconds. For a PortScan, we shall take as $N_i^\Delta(t)$ the number of different requested destination ports of the destination IP address $i$ in the $t$th sub-interval of size $\Delta$ seconds and for a NetScan, it will be the number of different requested destination IP addresses by the source IP address $i$. For Scan attacks, the source address is reliable (*spoofing* cannot be used) since the attackers wish to receive packets responding to their requests.

Our goal is now to provide algorithms for detecting change-points in the time series $(N_i^\Delta(t))_{t\geq 1}$ for each $i \in \{1,\ldots,D\}$ under computational constraints which make it possible to process the data on the fly, even for a high dimension $D$. For instance, in the case of TCP/SYN flooding, $D$ is the number of destination IP addresses appearing in the raw data and can be huge, up to several thousand addresses within a minute and several million within a few days.

For convenience, in the following we will drop the superscript $\Delta$ in the notation $N_i^\Delta(t)$.

## 3. Description of the methods.

We shall only consider batch methods in the following. The traffic is analyzed in successive observation windows, each having a duration of $P \times \Delta$ seconds, for some fixed integer $P$. A decision concerning the presence of potentially attacked IP addresses is made at the end of each observation window and we also identify the IP addresses involved. The value of $D$ then corresponds to the number of different $i$ encountered in the observation window of time length $P \times \Delta$ seconds.

A crude solution for detecting change-points in the time series $(N_i(t))_{1\leq t\leq P}$ would be to apply a change-point detection test to each time series $(N_i(t))_{1\leq t\leq P}$ for all $i \in \{1,\ldots,D\}$. Since $D$ can be huge even in a given observation window (see Figure 3 in Section 4.1), we are faced in practice with massive data streams leading to the construction and the analysis of several thousands of time series even for short observation periods (around 1 minute). To overcome this difficulty, a data reduction stage must precede the change-point detection step.

In the following, we propose two different data reduction mechanisms: record filtering and random aggregation (sketches). In short, the record filtering will select the heavy-hitters among the flows involved and process them, while the random aggregation will construct and process several (randomly chosen) linear combinations of all the flows. Random aggregation is currently the dimension reduction mechanism which is the most used in the network security community. Nevertheless, we believe that for the purpose of change-point detection, in particular, if the change-points involve a massive increase, record filtering would be a more efficient alternative. At first glance, random aggregation has the advantage of processing all the data. However, heavy-hitters are mixed with many other flows, which may smooth the change-points and result in poor detection. On the other hand, heavy-hitters are selected with high probability in record filtering and their change-points are more likely to be detected. To support this idea, we derive a toy problem related to this question in Section 6 and perform some numerical experiments on synthetic data in Section 7.

As for the change-point detection step, we propose using nonparametric tests based on $U$-statistics which do not require any prior information concerning the distribution of the time series constructed after the dimension reduction step. Such an approach is of particular interest for dealing with Internet traffic data, for which there is a lack of commonly accepted parametric models.

In the following we shall refer to record filtering followed by a nonparametric change-point detection test as the *TopRank* algorithm and when the record filtering stage is replaced by random aggregation, the *HashRank* algorithm. Both algorithms are further described below.

3.1. *The* TopRank *method.* The *TopRank* algorithm can be split into three steps described hereafter. Note that the following processing is performed in each observation window of length $P \times \Delta$ seconds and that all the stored data is erased at the end of each observation window.

STEP 1 (*Record filtering*). The main dimension reduction takes place in this step. In each sub-interval indexed by $t \in \{1, \ldots, P\}$ of duration $\Delta$ seconds of the observation window, we keep the indices $i$ of the $M$ largest $N_i(t)$ and label them as $i_1(t), \ldots, i_M(t)$ in such a way that $N_{i_1(t)}(t) \geq N_{i_2(t)}(t) \geq \cdots \geq N_{i_{M(t)}}(t)$. In the following we shall denote by $\mathcal{T}_M(t)$ the following set:

$$\mathcal{T}_M(t) = \{i_1(t), \ldots, i_M(t)\}.$$

In other words, for all $t \in \{1, \ldots, P\}$,

$$\#\mathcal{T}_M(t) = M \text{ and } \forall i \in \mathcal{T}_M(t) \text{ and } \forall j \notin \mathcal{T}_M(t), \qquad N_i(t) \geq N_j(t).$$

We stress that, in order to perform the following steps, we only need to store the variables $\{N_i(t), i \in \mathcal{T}_M(t), t = 1, \ldots, P\}$.

STEP 2 (*Creation of censored time series*). In this stage we shall build censored time series to be analyzed in the last step. For a given index $i$, the corresponding time series is censored since it is possible that for a given $t$ in $\{1, \ldots, P\}$, $i$ does not belong to the set $\mathcal{T}_M(t)$. In this case its value $N_i(t)$ is not available and is censored using the upper bound $N_{i_{M(t)}}(t) = \min_{i \in \mathcal{T}_M(t)} N_i(t)$. More formally, the censored time series is defined by

(1) $$(X_i(t), \delta_i(t))_{1 \leq t \leq P},$$

where, for each $t \in \{1, \dots, P\}$,

$$X_i(t) = \begin{cases} N_i(t), & \text{if } i \in \mathcal{T}_M(t), \\ \min_{j \in \mathcal{T}_M(t)} N_j(t), & \text{otherwise}, \end{cases}$$

$$\delta_i(t) = \begin{cases} 1, & \text{if } i \in \mathcal{T}_M(t), \\ 0, & \text{otherwise}. \end{cases}$$

The value of $\delta_i(t)$ tells us if the corresponding value $X_i(t)$ has been censored or not. Observe that, by definition, $\delta_i(t) = 1$ implies $X_i(t) = N_i(t)$ and $\delta_i(t) = 0$ implies $X_i(t) \geq N_i(t)$.

In practice, the censored time series are only built for indices $i$ selected in the first step, that is, $i \in \bigcup_{t=1}^{P} \mathcal{T}_M(t)$. However, many such time series will be highly censored. Hence, we propose an additional dimension reduction which consists in considering only the indices

$$i \in \bigcup_{t=1}^{P} \mathcal{T}_{M'}(t),$$

where $M' \in \{1, \dots, M\}$ is a chosen parameter.

STEP 3 (*Change-point detection test*). In Gombay and Liu (2000) a nonparametric statistical change-point detection method is proposed to analyze censored data, as well as a way of computing its $p$-values. It is a nonparametric rank test using a score function (denoted by $A$ in the following) which was first introduced by Gehan (1965) and Mantel (1967) in their generalization of Wilcoxon's rank test for censored data. We apply this test to each time series created in Step 2. Note that each of these time series is removed when the analysis in a given observation window is complete. With such an approach, up to $M' \times P$ time series of length $P$ are processed in each observation window of time length $P \times \Delta$ seconds.

Let us now further describe the statistical test that we perform. This procedure aims at testing from the observations $(X_i(t), \delta_i(t))_{1 \leq t \leq P}$ built in the previous step if a change occurred in the time series $(N_i(t))_{1 \leq t \leq P}$ for a given $i$. More precisely, if we drop the subscript $i$ for convenience in the description of the test, the tested hypotheses are as follows:

($H_0$): "$\{N(t)\}_{1 \leq t \leq P}$ are independent and identically distributed random variables."

($H_1$): "There exists some $r$ such that $(N(1), \dots, N(r))$ and $(N(r+1), \dots, N(P))$ have a different distribution."

Let us now describe the test statistic that we use. For each $s, t \in \{1, \dots, P\}$, we define the following:

- $A_{s,t} = \mathbb{1}(X(s) > X(t), \delta(s) = 1) - \mathbb{1}(X(s) < X(t), \delta(t) = 1)$, where $\mathbb{1}(E) = 1$ in the event $E$ and $0$ in its complementary set,
- $U_s = \sum_{t=1}^{P} A_{s,t}, s = 1, \ldots, P$,
- $S_t = (\sum_{s=1}^{t} U_s)/(\sum_{s=1}^{P} U_s^2)^{1/2}, t = 1, \ldots, P$.

We shall use

$$W_P = \max_{1 \leq t \leq P} |S_t| \tag{2}$$

as a test statistic. Since, under $(H_0)$ [see Gombay and Liu (2000)],

$$W_P \overset{d}{\longrightarrow} B^\star = \sup_{0 < t < 1} |B(t)| \qquad \text{as } P \to \infty,$$

where $\{B(t), t \in [0,1]\}$ denotes a Brownian bridge and $d$ the convergence in distribution, we shall take for the change-point detection test the following $p$-value: $Pval(W_P)$, where for all $b > 0$,

$$Pval(b) = \mathbb{P}(B^\star > b) = 2 \sum_{j=1}^{\infty} (-1)^{j-1} e^{-2j^2 b^2}, \qquad b > 0.$$

The last equality is given in Gombay and Liu (2000). These results can be proved by remarking that the numerator of $S_t$ is a $U$-statistic and the denominator is the normalization which ensures the convergence in distribution of $W_P$. All these results are proved in detail in Liu (1998) and Midodzi (2001), under specific assumptions. Then, for a given asymptotic level $\alpha \in (0,1)$, we reject $(H_0)$ when

$$Pval(W_P) < \alpha. \tag{3}$$

In the rejection case, the change-point instant is given by

$$\hat{t}_P = \underset{1 \leq t \leq P}{\mathrm{Argmax}} |S_t|.$$

3.2. *The* HashRank *method.* In this approach we propose another method to perform the data reduction stage based on hash functions. As explained previously, the following processing is performed in each observation window of length $P \times \Delta$ seconds and all the stored data is erased at the end of each observation window.

STEP 1 (*Filtering using hash functions*). Let us define a hash table with $L$ rows and $K$ columns. We consider $L$ hash functions $h_\ell, \ell = 1, \ldots, L$, each taking its value in $\{1, \ldots, K\}$. To each entry $(\ell, k) \in \{1, \ldots, L\} \times \{1, \ldots, K\}$ of the hash table is associated the list $\mathcal{L}_{\ell,k}$ of the indices $i$ which are hashed into this entry and a time series $(X_{\ell,k}(t))_{1 \leq t \leq P}$ defined by

$$X_{\ell,k}(t) = \sum_i \mathbb{1}(h_\ell(i) = k) N_i(t), \qquad t = 1, \ldots, P.$$

STEP 2 (*Change-point detection test*). We perform a statistical change-point detection test on each of the time series previously obtained. Let us denote by $\mathcal{C}$ the set of cells $(\ell, k), \ell \in \{1, \ldots, L\}, k \in \{1, \ldots, K\}$, of the hash table in which a change occurred. The test statistic that we use is the same as the one explained in Section 3.1 (nonparametric rank tests), except that the terms related to the censorship are removed.

STEP 3 (*Inversion of the hash table*). Assuming that a time series $(N_i(t))_{1 \leq t \leq P}$ which contains a change-point will yield a change-point in the $L$ time series

$$(X_{\ell, h_\ell(i)}(t))_{1 \leq t \leq P}, \qquad \ell = 1, \ldots, L,$$

we classify $i$ as an anomaly if a change-point has occurred in the time series associated to the cell $(\ell, h_\ell(i))$ for all $\ell \in \{1, \ldots, L\}$. In practice, the set of such indices is obtained from the lists $\mathcal{L}_{\ell,k}$ and the set $\mathcal{C}$ defined in the previous steps as follows:

$$\bigcap_{1 \leq \ell \leq L} \mathcal{L}_\ell^C, \qquad \text{where } \mathcal{L}_\ell^C = \bigcup_{k:(\ell,k)\in\mathcal{C}} \mathcal{L}_{\ell,k}.$$

Several types of hash functions can be considered in the first step, but we shall only focus on random hashing because of its low computational load.

We shall use 4-universal hashing functions for which a fast implementation is described in Thorup and Zhang (2004). Such functions are used since they are known to ensure a small number of collisions (a collision occurs when the $L$ different hashing functions have the same output when applied to two different indices $i$). Collisions may yield a false alarm during Step 3 described above. More precisely, following Abry, Borgnat and Dewaele (2007), we shall take as hash functions the following $(h_\ell)_{\ell=1,\ldots,L}$:

$$(4) \qquad h_\ell(x) = 1 + \left(\sum_{j=0}^{3} a_{j,\ell}\, x^k \bmod p\right) \bmod K,$$

where $x$ is a 32-bit integer (making this method particularly well suited for the hashing of IP addresses), and $p$ is the Mersenne prime number $2^{61} - 1$. As for the $(a_{j,\ell})$'s, they are picked randomly in $\{0, \ldots, p-1\}$ with independent outputs from one hash function to another.

## 4. Application to real data with attacks of SYN flooding type.

In this section we give the results of the *TopRank* algorithm when it is applied to some real Internet traffic provided by France-Télécom within the framework of the ANR-RNRT OSCAR project and we give some hints about the choice of the different parameters involved.

This data corresponds to a recording of 118 minutes of ADSL (Asymmetric Digital Subscriber Line) and *Peer-to-Peer* (P2P) traffic to which some TCP/SYN flooding type attacks have been added. Figure 1 (left) displays the total number of TCP packets received each second by the different requested IP addresses. The number of TCP/SYN packets received by the four attacked destination IP addresses are displayed on the right in Figure 1. As we can see in this figure, the first attack occurs at around 2000 seconds, the second at around 4000 seconds, the third at around 6000 seconds and the last one at around 6500 seconds.

From Figure 1, we can see that we are faced with massive data streams and that the attacks are completely hidden in TCP traffic and thus difficult to detect.

**4.1. *Choice of parameters.*** As described above, Step 1 (the record filtering step) and Step 2 (the creation of the censored time series) of the *TopRank* algorithm rely on several parameters ($P, \Delta, M$ for Step 1 and $M'$ for Step 2). As for Step 3 (the change-point detection test), it does not require the tuning of any parameters. The main objective of the first two steps is to cope with high dimensionality and the requirements of real time implementation. The choice of the parameters must satisfy these requirements as well as a reasonable average detection delay and a relevant selection of the time series to be processed in the subsequent detection step. In the following, we give some guidance regarding parameter selection in the context of Network anomaly detection.

Since a decision concerning the presence of attacks is made at the end of each observation window, the maximal detection delay is given by the time length of the observation window, that is, $P \times \Delta$ seconds. The average detection delay is then given by $P \times \Delta/2$ seconds (typically about 30 seconds in the context of Network anomaly detection). Once the average detection delay has been chosen, one should choose $P$ as large as possible under the constraints of real time data processing for a given maximal number of tests to perform within the observation window. Indeed, a large $P$ ensures a better statistical consistency. On the other hand, the test statistic $W_P$ in (2) has an $O(P^2)$ computational complexity. Given the computational limits of a standard computer, we chose $P = 60$ in order to allow up to $10^3$ time series to be processed within a 1 minute observation window.

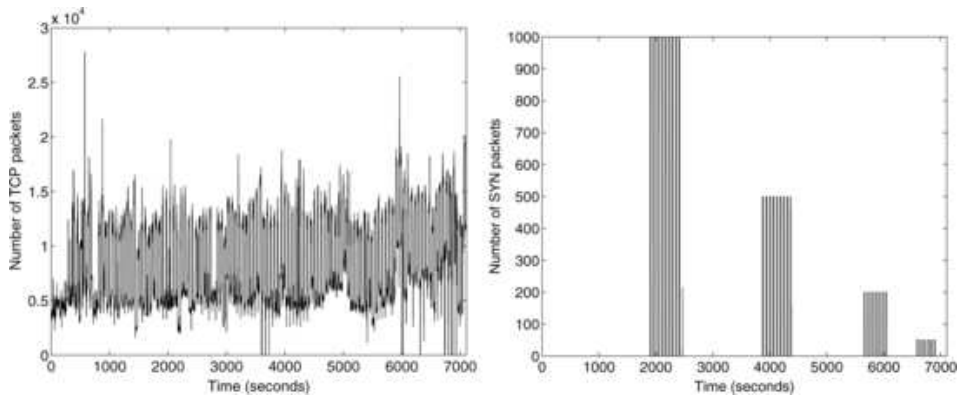

FIG. 1. *Number of TCP packets exchanged and number of TCP/SYN packets received by the 4 attacked IP addresses.*

Let us now explain the choice of $M$, which sets the censorship level. Indeed, the number of TCP/SYN packets received by the $M$th most requested machine corresponds to a threshold above which an IP address may appear as potentially attacked. From the data, we remark that 99% of the observed values of this threshold are at most 10 when $M = 10$ and at most 5 for $M = 20$. In the applications to real data, we chose $M = 10$ to allow us to capture flows with significantly high traffic rates (10 packets per second) while ensuring a low cost in terms of memory storage. Recall that a $M \times P$ data table $(X_i(t), \delta_i(t))_{1 \le t \le P, 1 \le i \le M}$ has to be stored during the first and second step.

We now comment on the choice of the parameter $M'$ in $\{1, \ldots, M\}$. Taking $M' = 1$ means that we only analyze the IP addresses $i$ having, at least once in an observation window, the largest $N_i(t)$. Taking $M' = M$ means that the detection step is applied to the censored time series of all IP addresses $i$ which have been selected in the filtering step. Hence, increasing $M'$ increases the number of analyzed censored time series. Figure 2 displays the number of time series which are actually built in Step 2 of *TopRank* after the filtering stage of Step 1 for different values of $M'$: $M' = 1$, $M' = 5$ and $M' = M = 10$ when we are looking for TCP/SYN flooding type attacks. Figure 3 displays the number of different destination IP addresses every minute of the traffic trace. Comparing Figure 2 with Figure 3, we see that Steps 1 and 2 of the *TopRank* algorithm appear to be necessary to provide an implementation feasible on the fly. Indeed, applying Step 3 to each IP address every minute would produce an excessive computational load. As for the statistical performance of the method with respect to the parameter $M'$, we shall see in the following section that the parameter $M'$ does not significantly change the results in terms of false alarm and detection rates.

4.2. *Performance of the method.* We now investigate the performance of the *TopRank* algorithm with the following parameters: $P = 60$, $\Delta = 1$ s, $M = 10$ and $M' = 1$. Using these parameters, the average detection delay is 30 seconds.

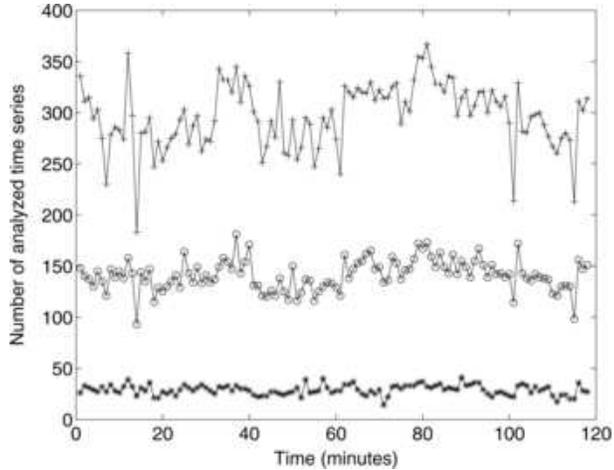

Fig. 2. *Number of analyzed time series every minute when $M' = 1$ ("*"), $M' = 5$ ("o") and $M' = M = 10$ ("+").*

4.2.1. *Statistical performance.* First, note that with the previous choice of parameters the attacked IP addresses have been identified when the upper bound of the $p$-value $\alpha$ introduced in Step 3 of *TopRank* is such that $\alpha \geq 2 \times 10^{-6}$.

Figure 4 displays the censored time series (Step 2 of *TopRank*) of the four attacked IP addresses. These censored time series are displayed in the first observation window in which the algorithm detects the anomaly. We also display with a vertical line the instant where the change is detected. Re-member that the uncensored time series of these 4 IP addresses are displayed in Figure 1 (right).

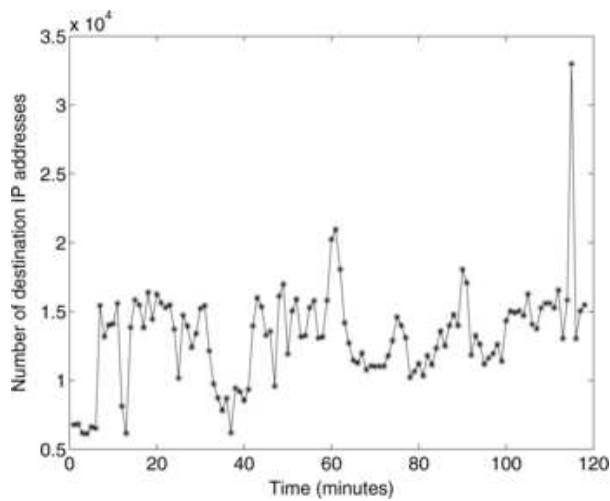

Fig. 3. *Number of destination IP addresses every minute.*

The *TopRank* part of Table 1 gives the smallest *p*-value above which the corresponding attack is detected, as well as the number of false alarms. The number of false alarms corresponds to the number of IP addresses for which an alarm is triggered but which are different from the attacked IP addresses. For instance, the first attack is detected if $\alpha \geq 10^{-8}$ [see (3)] and the associated number of false alarms is equal to 3. If $M' = 5$ or $M' = M = 10$, the results remain unchanged except for the third attack for which the number of false alarms equals 10 instead of 9.

In Table 1 we also give the results obtained from the same data with a method proposed by Siris and Papagalou (2004). This algorithm uses the CUSUM algorithm to look for a change-point in the time series corresponding to the sum of received SYN packets by all the destination IP addresses which have been requested. For each observation window of 60 seconds, an alarm is set off when the statistic $g_n$ defined in equation (6) of Siris and

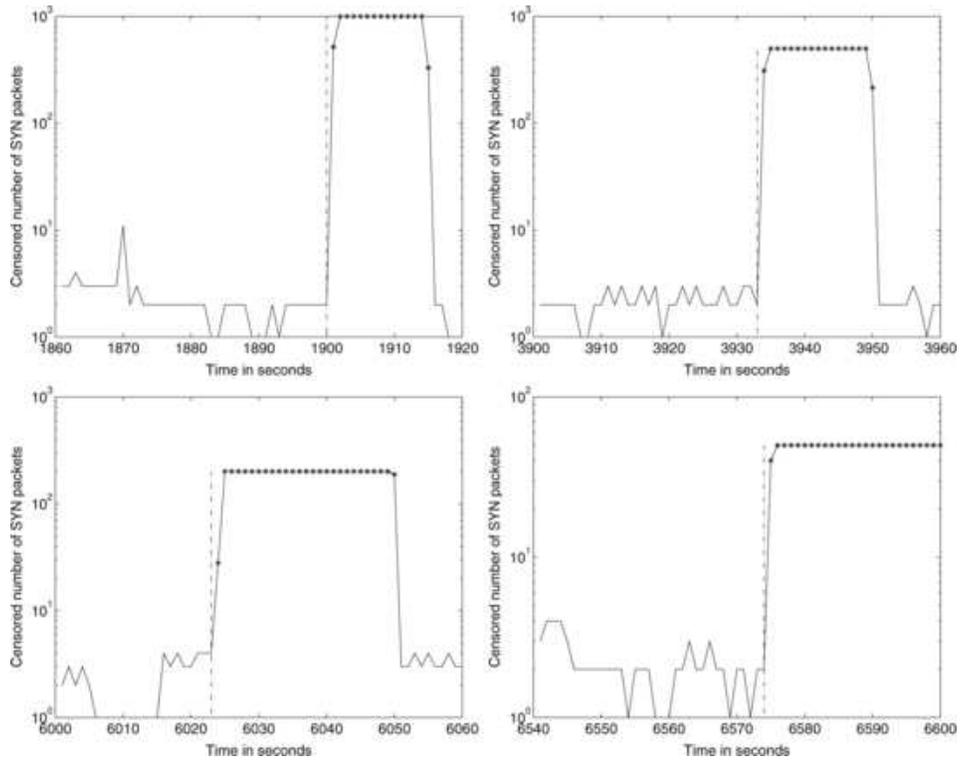

FIG. 4. *Censored time series of the 4 attacked IP addresses, where the vertical lines correspond to the detected change-point instants and the uncensored values are displayed with stars ("∗").*

TABLE 1

*Statistical performance for detecting the 4 successive SYN flooding attacks displayed on the right-hand side of Figure 1. The attacks consist in sending SYN packets to a given destination IP address. In the top row the intensity (number of SYN packets per second) of each attack is given. In the* TopRank *part of the table, the displayed p-values are the smallest ones that ensure the detection of the attack by the* TopRank *algorithm and below that the corresponding number of false alarms in the whole traffic trace is given. In the SP part of the table [SP is the method devised by Siris and Papagalou (2004)], the h row gives the smallest threshold values that ensure the detection of each attack. In the last row the corresponding number of false alarms is displayed*

|         | Number of SYN packets | 1000 | 500 | 200 | 50 |
|---------|----------------------|------|-----|-----|-----|
| *TopRank* | $p$-values | $10^{-8}$ | $10^{-10}$ | $2 \times 10^{-6}$ | $10^{-12}$ |
|         | Number of false alarms | 3 | 1 | 9 | 0 |
| SP | $h$ | 5 | 6.5 | 9.7 | 16.34 |
|    | Number of false alarms | 69 | 65 | 62 | 22 |

Papagalou (2004) is greater than a threshold $h$ at least once in the window. This quantity $g_n$ depends on two parameters $\alpha$ and $\beta$. We use the same values as Siris and Papagalou (2004), namely, $\alpha = 0.5$ and $\beta = 0.98$, to obtain the results displayed in Table 1. We observe that the *TopRank* algorithm allows us not only to retrieve the attacked destination IP addresses, but also seems to perform better in terms of false alarm rate. This suggests that aggregating traffic flows results in a poor detection of malicious flows, especially when the normal traffic is high. Indeed, a close look shows that the normal traffic is particularly high during the first attack, which explains why the corresponding threshold value is the lowest (and the false alarm rate the highest) although this attack is the most intense. We shall investigate the comparison between record filtering and aggregation filtering further, in Sections 6 and 7, in terms of information loss and detection performance respectively.

To compute the number of false alarms, we have considered that the attacked IP addresses were only those for which an attack was generated, but it is possible that the underlying ADSL and P2P traffic contains some other attacks. Figure 5 displays the censored time series of some IP addresses which were considered to be false alarms in the *TopRank* part of Table 1, as well as the time instant where a change was detected (vertical line). However, if we refer to their time series, these IP addresses could be considered as being attacked. Thus, the results shown in Table 1 have been computed in the most unfavorable way for the algorithms.

### 4.2.2. *Numerical performance.*
As we have seen, the *TopRank* algorithm seems to give satisfactory results from a statistical point of view. Moreover,

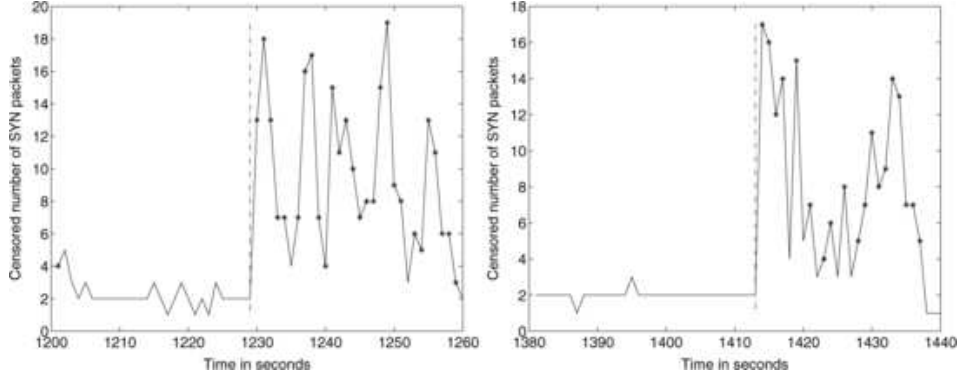

Fig. 5. *Censored time series for IP addresses considered as false alarms, where the vertical lines correspond to the detected change-point instants and the uncensored values are displayed with stars ("∗").*

with $M = 10$, $M' = 1$, and $P = 60$, applying the *TopRank* algorithm takes only 1 minute and 19 seconds to process the whole traffic trace of 118 minutes, when looking for TCP/SYN flooding type attacks with a computer having the following configuration: RAM 1 GB, CPU 3 GHz.

**5. Application to real data with several kinds of attacks.** In this section we shall show that the *TopRank* algorithm can not only detect SYN flooding type attacks but also any other kind of attacks, such as UDP flooding, PortScan and NetScan attacks. As we shall see in the following, the *TopRank* algorithm can detect, identify the anomaly and also provide the IP addresses involved.

The real data of this section has been provided by France-Télécom Internet Service Provider within the framework of the ANR-RNRT OSCAR project. It corresponds to a recording of 67 minutes of ADSL and P2P traffic to which some attacks of the following types, SYN flooding, UDP flooding, PortScan and NetScan, have been added.

The top of Figure 6 displays the total number of TCP packets, as well as the number of TCP packets received by the destination IP address attacked by a PortScan. The bottom of Figure 6 displays the total number of packets, as well as the number of packets sent by the source IP address generating a NetScan attack.

In Figure 6 we can see, as in Section 4, that the attacks are completely hidden in the total traffic and thus difficult to detect.

5.1. *Choice of parameters.* The previous data has been processed using the same parameters as those used in Section 4: $P = 60$, $\Delta = 1$ s, $M = 10$ and $M' = 1$. Note that the reduction stage (Step 1 of *TopRank*) is also necessary for the analysis of this dataset. Indeed, in Figure 7, which displays the number of different destination IP addresses each minute of the traffic trace, we can see that Step 3 cannot be applied to each IP address at every minute in a reasonable computational time.

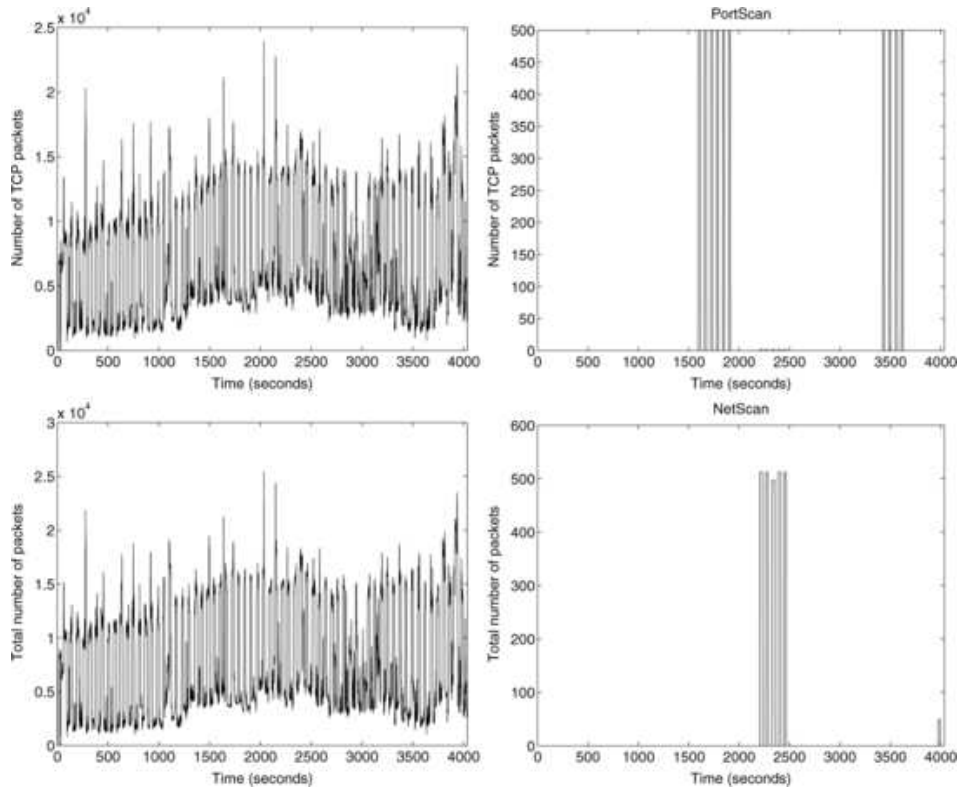

Fig. 6. *Number of exchanged packets for PortScan and NetScan attacks.*

### 5.2. *Performance of the method.*

5.2.1. *Statistical performance.* First, note that with the previous choice of parameters the attacked IP addresses have been identified when the upper bound of the $p$-value $\alpha$ introduced in Step 3 of *TopRank* is such that $\alpha \geq 10^{-11}$ for the PortScan, $\alpha \geq 10^{-6}$ for the UDP flooding, $\alpha \geq 0.0006$ for the SYN flooding and $\alpha \geq 0.04$ for the NetScan.

Figure 8 displays the censored time series (Step 2 of *TopRank*) of the attacked IP addresses in the case of SYN flooding, UDP flooding and PortScan, as well as the censored time series of the source IP address generating the attack in the case of NetScan. These time series are displayed in the first

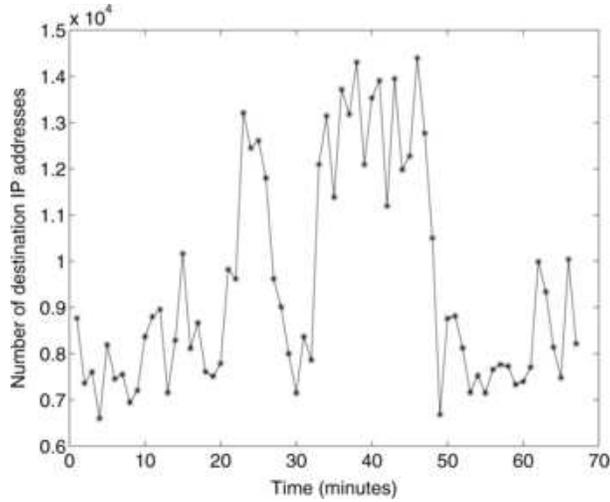

Fig. 7. *Number of destination IP addresses each minute.*

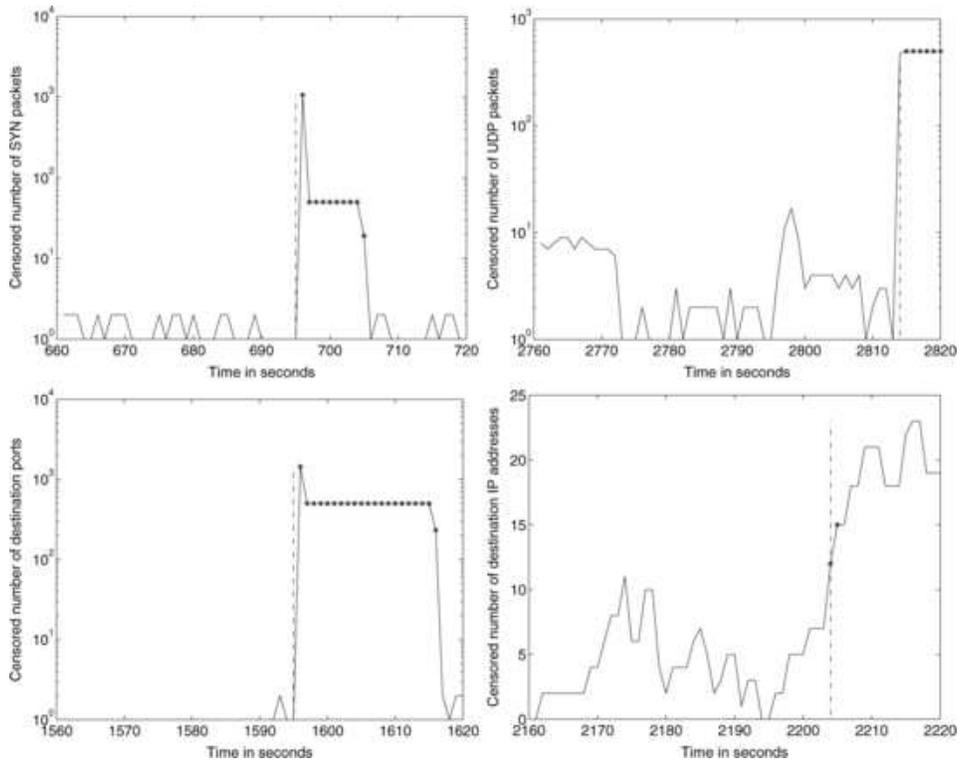

Fig. 8. *Censored time series of the attacked IP addresses, where the vertical lines correspond to the detected change-point instants and the uncensored values are displayed with stars ("∗").*

TABLE 2

*Statistical performance of the* TopRank *algorithm for detecting several types of attacks. In the second row are displayed the smallest p-values that ensure the detection of the PortScan, SYN flooding, UDP flooding and NetScan type attacks respectively. In the last row, the corresponding number of false alarms is given*

|                        | **PortScan** | **SYN flooding**   | **UDP flooding** | **NetScan** |
| ---------------------- | ------------ | ------------------ | ---------------- | ----------- |
| $p$-values             | $10^{-11}$   | $6 \times 10^{-4}$ | $10^{-6}$        | 0.04        |
| Number of false alarms | 3            | 127                | 136              | 378         |

observation window in which the algorithm detects the anomaly of the corresponding type. We also display with a vertical line the instant where the change is detected. The detection time delay is equal to around 1 minute for the SYN flooding, 5 seconds for the UDP flooding, 30 seconds for the PortScan and 20 seconds for the NetScan.

We can see from Figure 8 that the rank test for censored data described in Step 3 of *TopRank* can detect several types of changes: sudden increase, slow increase and several types of magnitude of changes.

Table 2 gives the smallest $p$-value above which the corresponding attack is detected, as well as the number of false alarms. The number of false alarms corresponds to the number of IP addresses for which an alarm is triggered by the *TopRank* algorithm but which are different from the attacked IP addresses. For instance, the PortScan attack is detected if $\alpha \geq 10^{-11}$ and the associated number of false alarms is equal to 3.

To compute the number of false alarms, we have considered, as previously, that the attacked IP addresses were only those for which an attack was generated, but it is possible that the background ADSL and P2P traffic contains some other attacks. Figure 9 displays the censored time series of some IP addresses which were considered to be false alarms in the computation of Table 2 in the case of the SYN flooding, UDP flooding and NetScan, as well as the time instant where a change was detected (vertical line).

Since these IP addresses could be considered as being attacked, the results of Table 2 are computed in the most unfavorable way for our algorithm. This is a major problem for comparing algorithms on massive data streams. The issue is that without properly labeled datasets, comparisons cannot be performed using standard methods such as ROC (Receiver Operating Characteristic) curves. A second difficulty is to distinguish the relative merits of the filtering step and those of the detection step. Finally, a third difficulty arises in the specific case of Network anomaly detection: Internet traffic is known to be bursty, with loads varying quite a lot over the day. Hence, the performance observed on a 2 hour dataset cannot be generalized without caution, especially for methods that rely on the choice of numerous parameters. The parameters of the filtering step can be interpreted in terms of dimension reduction rate and are thus easier to compare. This is much more difficult as far as the detection step is concerned, since the parameters often need to be adapted to the statistical characteristics of the traffic data being studied. The methods proposed by Krishnamurthy et al. (2003), Li et al. (2006), Salem, Vaton and Gravey (2007) and Abry, Borgnat and Dewaele (2007) all use as a dimension reduction technique random aggregation (sketches) which enables them to work on the fly. All these methods use either outlier detection, or CUSUM in their change detection step, with several parameters (in addition to the test threshold) that need to be fine-tuned. In practice, because of the high variability of the loads within one day, the choice of these parameters might require some expertise. In contrast, our detection step is purely nonparametric. Since it is based on rank statistics, it only requires choosing a threshold of the $p$-value under which an alarm is set off.

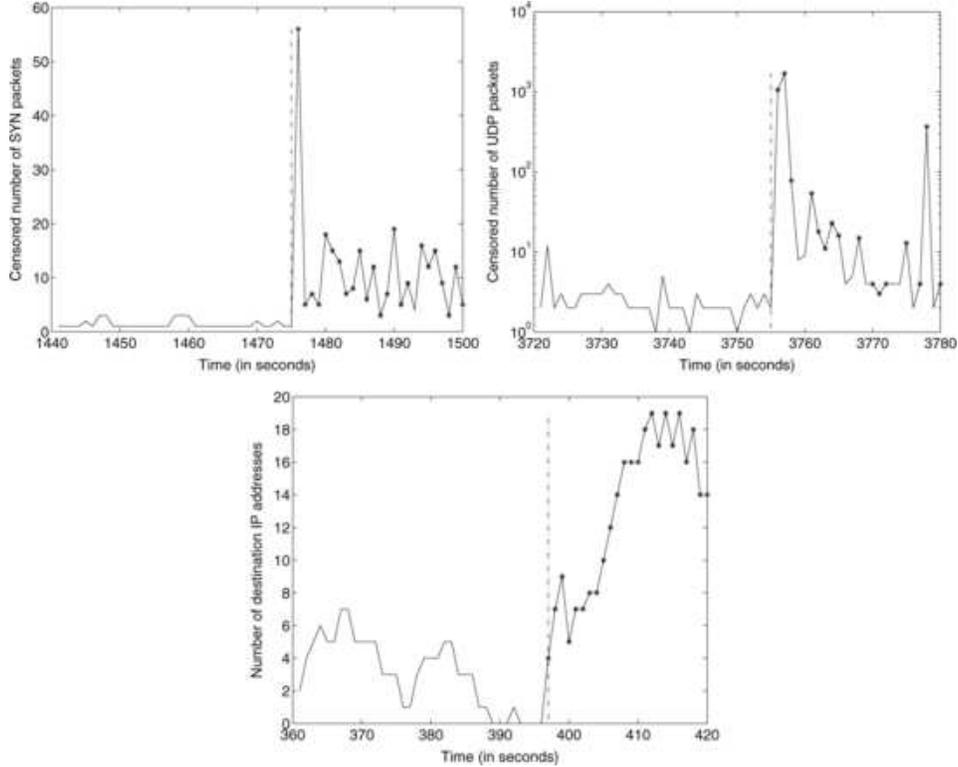

Fig. 9. *Censored time series for IP addresses considered as false alarm in the case of SYN and UDP flooding (top: left and right respectively) and in the case of NetScan (bottom). The vertical lines correspond to the detected change-point instants and the uncensored values are displayed with stars ("∗").*

In Section 7 we shall compare our filtering steps with random aggregation filtering on synthetic datasets which are completely labeled.

5.2.2. *Numerical performance.* As we have seen, this method seems to give satisfactory results from a statistical point of view. Moreover, with $M = 10$, $M' = 1$, and $P = 60$, applying the *TopRank* algorithm takes only 1 minute and 30 seconds to process the whole traffic trace of 67 minutes, when looking for SYN flooding, UDP flooding, PortScan and NetScan, with a computer having the following configuration: RAM 1 GB, CPU 3 GHz. This makes the implementation of the *TopRank* algorithm very realistic in detection systems processing data on the fly, even for very intense traffic data.

**6. A toy problem comparing record and aggregation methods.** In this section we briefly introduce and solve a toy problem to compare the effects of record filtering and random aggregation on data dimension reduction in a simplified framework. We have considered these two data reduction methods in the first steps of the *TopRank* and *HashRank* algorithms, which aim at detecting an increase of one (or several) components of a $D$-dimensional time series for a large $D$. For detecting such anomalies, one has to be able to infer from the reduced data, at each time instant, that one of the components is indeed large. We consider this inference problem in a case where all the components have the same known distribution, except possibly one, whose distribution is obtained by multiplying by a scale factor $\theta^{-1} \geq 1$. Hence, we are in a standard problem of estimating a scalar parameter $\theta$. To compare the theoretical merits of the two data reduction methods, we simply evaluate the Fisher information related to the two different observations.

More precisely, let $X_1, \ldots, X_D$ be $D$ independent random variables with positive values such that, among them, $D - 1$ admit the same density $p(x)$ and one admits the density $\theta p(\theta x)$ with $\theta \in (0, 1]$. We shall compare the Fisher information associated with the parameter $\theta$ for two different observations:

1. The observation obtained by data dimension reduction using simple record filtering:

$$(5) \qquad\qquad Y_D = \max_{k=1,\ldots,D}(X_k).$$

2. The observation obtained by data dimension reduction using simple random aggregation filtering:

$$(6) \qquad\qquad Z_D = \sum_{k=1}^{D} X_k.$$

The distributions of $Y_D$ and $Z_D$ are both parametrized by $\theta$. We will denote the corresponding densities by $f_{D,\theta}$ and $g_{D,\theta}$ respectively, and the corresponding Fisher information quantities [see Bickel and Doksum (1976)] by

$$I_D(\theta) = \mathrm{Var}(\partial_\theta \log f_{D,\theta}(Y_D)) \quad \text{and} \quad J_D(\theta) = \mathrm{Var}(\partial_\theta \log g_{D,\theta}(Z_D)).$$

Suppose that $p$ is 2 times continuously differentiable, has its support in $[0, 1]$ and that

$$\mathbb{E}\left[\left(\frac{\dot{p}}{p}(X)\right)^2\right] < \infty,$$

where $\dot{p}$ denotes the derivative of $p$ and $X$ a random variable with density $p$. We claim that, for any $\theta \in (0, 1)$, as $D \to \infty$,

$$(7) \quad I_D(\theta) \to \frac{\text{Var}((\theta \vee X)(\dot{p}/p)(\theta \vee X))}{\theta^2} \quad \text{and} \quad J_D(\theta) \sim D^{-1} \frac{(\mathbb{E}[X])^2}{\theta^4 \text{Var}(X)},$$

where $a \vee b = \max(a, b)$.

REMARK 1. Since $I_D(\theta) \asymp 1$ and $J_D(\theta) \asymp D^{-1}$, this result clearly shows that, for large $D$, record filtering is advantageous from an information point of view in comparison with aggregation filtering. However, the fact that $p$ has compact support is crucial in this result. We believe that $I_D$ would have different asymptotic behavior if other assumptions were made on the tail distribution of $X$. However, we leave this question open for future work.

Let us give some simple arguments to support our claims. It is easy to show that, as $D \to \infty$,

$$Y_D \xrightarrow{d} 1 \vee (X/\theta).$$

We admit that this convergence in distribution can be strengthened to establish that the Fisher information $I_D$ tends to the one associated to the limit distribution. The limiting distribution has density $t \mapsto \theta p(\theta t) / \int_\theta^1 p(u) \, du$ with a support in $[0, 1/\theta]$ and the corresponding Fisher information is equal to $\text{Var}((1 \vee X/\theta)\frac{\dot{p}}{p}(\theta \vee X))$. Hence, the left-hand side of (7).

Denote $\sigma^2 = \text{Var}(X)$ and $\mu = \mathbb{E}[X]$ and $\tilde{Z}_D = (\sqrt{D-1}\sigma)^{-1}(Z_D - (D-1)\mu)$. Observe that the Fisher information associated with the observation $\tilde{Z}_D$ is the same as for $Z_D$. The centering and normalization is chosen so that the sum of the $(D-1)$ i.i.d. $X_i$'s in $Z_D$ is approximately $\mathcal{N}(0, 1)$. The remaining $X_i$ is then divided by $\sqrt{D-1}$ and thus is negligible as $D \to \infty$,

$$\tilde{Z}_D \xrightarrow{d} U \qquad \text{with } U \sim \mathcal{N}(0, 1).$$

In contrast with the previous case, as $D \to \infty$, the asymptotic distribution does not depend on $\theta$. This is why $J_D(\theta) \to 0$, as $D \to \infty$. To obtain an equivalent as in (7), we use

$$\tilde{Z}_D = U + (\theta\sqrt{D-1}\sigma)^{-1}X$$

to approximate the distribution of $\tilde{Z}_D$, which is more precise than the limit distribution. The corresponding density is given by the convolution

$$\hat{p}_\theta(x) = \sqrt{D-1}\sigma\theta \int_{t=-\infty}^{\infty} g(x-t)p(\sqrt{D-1}\sigma\theta t)\,dt,$$

where $g$ denotes the density of $U$. It follows that

$$(8)\qquad \partial_\theta \log \hat{p}_\theta(x) = \theta^{-1} - \frac{\int g(x-t)q(\sqrt{D-1}\sigma\theta t)}{\theta \int g(x-t)p(\sqrt{D-1}\sigma\theta t)},$$

where $q(t) = -t\dot{p}(t)$. Denote by $\dot{g}$ the derivative of $g$. Using a Taylor expansion of $g(x-t)$ around $x$, one has for a kernel function $k$ such that $\int k = 1$, as $b \to 0$,

$$\int g(x-t)k(t/b)\,dt = b\Big(g(x) - b\dot{g}(x)\int tk(t)\,dt + O(b^2)\Big).$$

Observing that $\int q = \int p = 1$, we may apply this approximation to the convolutions appearing in the ratio displayed in (8) with $b = (\sqrt{D-1}\sigma\theta)^{-1}$ and $k = q$ and $p$ successively. Since $\int tq(t)\,dt = 2\int tp(t)\,dt = 2\mu$, we get, as $D \to \infty$,

$$\partial_\theta \log \hat{p}_\theta(x) = \theta^{-1}\left[1 - \frac{g(x) - 2\mu b\dot{g}(x) + O(b^2)}{g(x) - \mu b\dot{g}(x) + O(b^2)}\right]$$
$$= \theta^{-1}\mu b\dot{g}(x)/g(x) + O(b^2).$$

Using this approximation, since $\dot{g}(x)/g(x) = -x$ [recall that $g$ is the density of the $\mathcal{N}(0,1)$] and $\mathrm{Var}(\tilde{Z}_D) \sim 1$, we get, as $D \to \infty$, $\mathrm{Var}(\partial_\theta \log \hat{p}_\theta(\tilde{Z}_D)) \sim D^{-1}\frac{\mu^2}{\theta^4\sigma^2}$. Admitting that this provides a good approximation of the Fisher information of $Z_D$, we get the right-hand side of (7).

**7. Application to synthetic data.** In Sections 4 and 5 the traffic traces are not completely labeled. We only have some labeled anomalies, those generated in the experiments, but many other anomalies, present in the background ADSL and P2P traffic, are not labeled. Indeed, the amount of data is too large and all the destination IP addresses cannot be thoroughly analyzed to make a diagnostic of an anomaly for each of them. Here, we generate synthetic high-dimensional data corresponding to 1 minute of Internet traffic and containing 1 anomaly. Using Monte Carlo simulations, ROC curves are obtained for the different detection algorithms, and then compared.

7.1. *Description of the synthetic data.* We explain, in the following, how the number of SYN packets received by the different IP addresses involved in a given observation window is synthesized in each sub-interval of time length $\Delta$. For a given IP address, we propose modeling the SYN packet traffic using a Poisson point process with a given intensity expressed as a number of SYN packets received per second. In Network applications, each IP address receives a very different amount of traffic. Hence, we shall use different intensities from an IP address to another. To take into account this diversity, we propose using the realizations of a Pareto distribution for the parameters of the different intensities so that a lot of machines receive a small number of SYN packets while a few receive a lot.

Let us now further describe the framework of our numerical experiments. We first randomly generate a sequence $(\theta_i)_{i=1,\ldots,D}$ of intensities for $D = 1000$ IP addresses with the Pareto distribution having the following density: $\gamma\alpha/(1 + \gamma x)^{1+\alpha}$, when $x > 0$, with $\alpha = 2.5$ and $\gamma = 0.72$, which roughly corresponds to what we observed in the real traffic traces at our disposal. Then, we generate the $D$ time series of length $P = 60$ corresponding to the number of packets received by each IP address in the 60 sub-intervals of the observation window, which consists of i.i.d. Poisson random variables, except for 1 IP address for which a change-point is introduced. Let us denote by $(Y_{i,j})$, where $1 \leq i \leq D$ and $1 \leq j \leq P$, the data thus obtained and by $i_0$ the IP address whose time series contains a change-point at time index $j_0 \in \{1, \ldots, P\}$:

$$(Y_{i_0,j})_{1 \leq j \leq j_0} \overset{\text{i.i.d.}}{\sim} \mathcal{P}\text{oisson}(\theta_{i_0}) \quad \text{and} \quad (Y_{i_0,j})_{j_0 < j \leq P} \overset{\text{i.i.d.}}{\sim} \mathcal{P}\text{oisson}(\eta\theta_{i_0}),$$

where the parameter $\eta$ is a positive coefficient. The remaining time series are generated as follows:

$$\forall i \neq i_0, \qquad (Y_{i,j})_{1 \leq j \leq P} \overset{\text{i.i.d.}}{\sim} \mathcal{P}\text{oisson}(\theta_i).$$

In the following, we shall take $D = 1000$, $j_0 = 35$, $P = 60$ and $\eta = 2$ or $\eta = 7$. A sensible choice of $i_0$ is made by first sorting the IP addresses $i \in \{1, \ldots, D\}$ so that the parameters $\theta_i$ are in a decreasing order, $\theta_1 \geq \theta_2 \geq \cdots \geq \theta_D$. The choice of $i_0$ is thus related to the difficulty of the detection: the smaller $i_0$, the easier the detection. For $\eta = 2$, we chose $i_0 = 100$ (10th percentile intensity) and for $\eta = 7$, we chose $i_0 = 500$ (median intensity), for which we have $\theta_{100} \approx 2$ and $\theta_{500} \approx 0.4$. See Figure 10, where the $(\theta_i)_{1 \leq i \leq D}$ are displayed.

Figure 11 displays the time series which have been synthesized using the previously described model corresponding to $i = 1$, $i = i_0$ and $i = 10$. The case where $i_0 = 100$ and $\eta = 2$ is displayed on the left and the case where $i_0 = 500$ and $\eta = 7$ on the right.

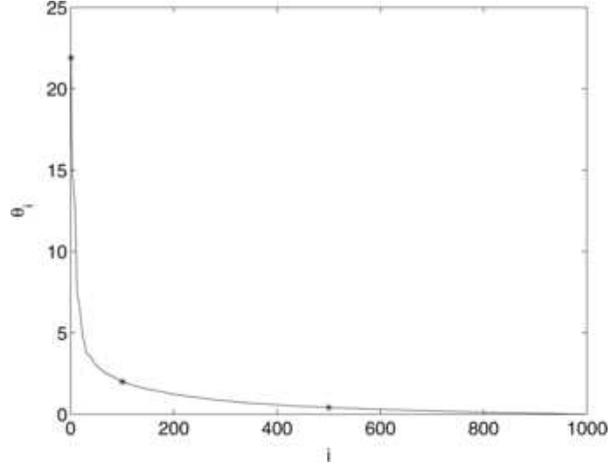

Fig. 10. *An example of parameters* $(\theta_i)_{1 \leq i \leq D}$. *The stars "*∗*" correspond to* $\theta_1$, $\theta_{100}$ *and* $\theta_{500}$.

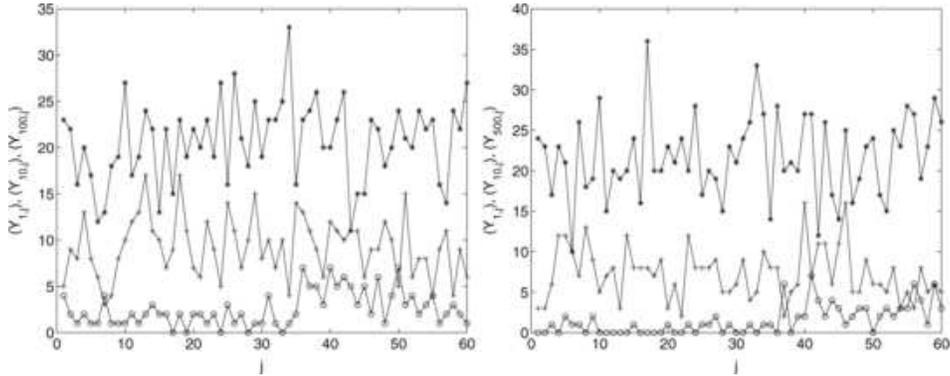

Fig. 11. *Some synthesized time series* $(Y_{i,j})_{1 \leq j \leq 60}$. *Left:* $i = 1$ *("*∗*"),* $i = 10$ *("+"),* $i = i_0 = 100$ *with* $\eta = 2$ *("o"). Right:* $i = 1$ *("*∗*"),* $i = 10$ *("+"),* $i = i_0 = 500$ *with* $\eta = 7$ *("o").*

**7.2.** *ROC curves of the* TopRank *and* HashRank *algorithms.* Now, we propose comparing the two approaches presented in Section 3 by computing their false-alarm and detection rates, which leads to the plot of the ROC curves displayed hereafter. More precisely, the $x$-axis and $y$-axis of the ROC curves that we consider correspond to the false-alarm rate (the number of false anomalies divided by $D - 1$, the number of time series without anomalies) and the detection rate (1 if the anomaly is detected and 0 otherwise), both averaged over 100 successive Monte Carlo experiments. Each point of the curve is obtained for a given threshold *Pthresh* below which a $p$-value is classified as an anomaly.

To make this comparison as fair as possible, the *TopRank* algorithm has been modified in order to address a number of time series in the change-point detection step, equal to the number addressed by the *HashRank* algorithm. Using this convention, we ensure an equivalent computational load for both algorithms, since, in both cases, the detection step is similar (nonparametric rank test).

In the case of the *HashRank* algorithm, the number of time series to which a detection test is applied is equal to $L \times K$ with $L = 8$ and $K = 17$, these two values ensuring a small number of collisions. In our experiments, the smallest value of $M$ for which $8 \times 17 = 136$ time series can be constructed in the *TopRank* algorithm is $M = 50$. For this $M$, we construct the 136 first time series encountered as the index $i$ goes along the list $i_1(1), i_1(2), \ldots, i_1(60), i_2(1), i_2(2), \ldots, i_2(60), i_3(1), \ldots$, where the $i_k(t)$ are defined in Step 1 of the *TopRank* algorithm; see Section 3.1.

In Figure 12 are displayed the ROC curves of the *TopRank* and *HashRank* algorithms when $i_0 = 500$ and $\eta = 7$ (left) and when $i_0 = 100$ and $\eta = 2$ (right). We also plot the ROC curve $y = x$ corresponding to the random classifier and the ROC curve of the *Comprehensive Rank* test (which consists in applying the nonparametric rank test to each time series, i.e., without any reduction step). Remember that such an approach is not feasible in practice for anomaly detection on the fly. It is displayed here to highlight the loss resulting from the data reduction steps.

Let us comment briefly on these results. It appears that the *TopRank* algorithm outperforms the *HashRank* algorithm except in the first case for high false alarm rates. Indeed, because of the record filtering, it is possible that the time series indexed by $i_0$ is not analyzed in the detection step and thus not detected, no matter how close to 1 the threshold *Pthresh* is chosen. This is why, on the left in Figure 12, the detection rate is bounded from above by 0.88. This does not occur on the right because $\eta \times \theta_{i_0}$ is large enough so that the time series indexed by $i_0$ is analyzed with a probability almost equal to 1. As a result, the performance of the *TopRank* algorithm is similar to the *Comprehensive Rank* test, that is, as if no reduction step were applied. This indicates that the censorship does not affect the sensitivity of the detection.

Figure 13 shows that the *TopRank* algorithm seems to reach a detection rate very close to 1 for a false alarm rate of around 0.04, while the *Comprehensive Rank* test attains this detection for a false alarm rate of around 0.138. This can be explained as follows: some time series falsely detected as an anomaly by a nonparametric rank test are not selected by the record filtering step, which diminishes the number of false alarms. More generally, for the *TopRank* algorithm, the number of alarms is bounded by the number of analyzed time series [$L \times K = 136$, which corresponds to a false alarm rate $\{(L \times K) - 1\}/(D - 1) \approx 0.136]$. This is not the case for the *HashRank* algorithm, although only $L \times K$ time series are analyzed. Indeed, the analyzed time series are random aggregations of all the raw time series, and all the IP addresses can be retrieved in the inversion step.

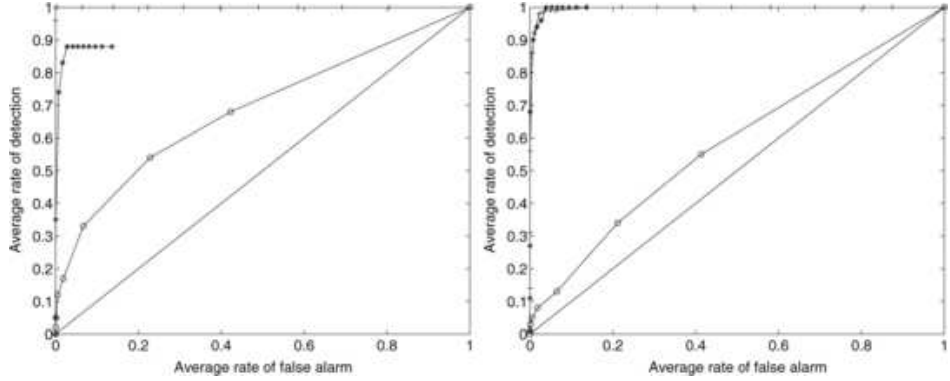

Fig. 12.  *ROC curves using* TopRank *("∗"),* HashRank *("o"), random classifier (plain) and the* Comprehensive Rank *test ("+"). The anomaly is simulated with, on the left, $i_0 = 500$, $\eta = 7$, on the right, $i_0 = 100$, $\eta = 2$.*

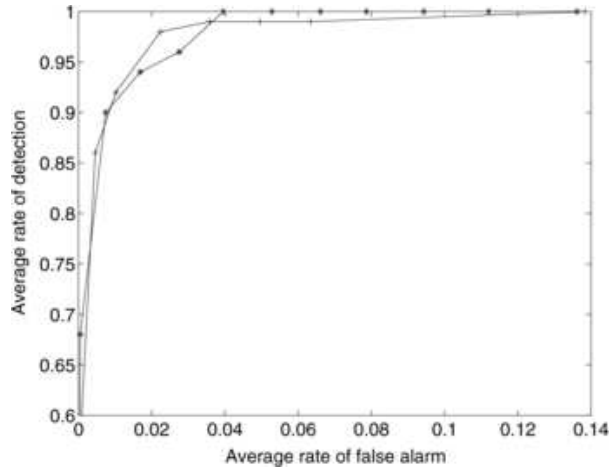

Fig. 13.  *Zoom on the top-left part of the graph on the right in Figure 12.*

Finally, it is interesting to observe that, in contrast to the *TopRank* algorithm, the *HashRank* algorithm has better performance in the first situation when $i_0 = 500$ and $\eta = 7$ than in the second, when $i_0 = 100$, $\eta = 2$. This is because in the first case the jump at the change-point is higher, $(\eta - 1)\theta_{i_0} \approx 6 \times 0.4 = 2.4$ in comparison to $(\eta - 1)\theta_{i_0} \approx 2$.

**8. Conclusion.**   In this paper we propose and compare two anomaly detection methods for identifying DoS attacks in Internet traffic: the *TopRank* algorithm based on a record filtering technique followed by a nonparametric rank test for censored data and the *HashRank* algorithm based on random aggregation followed by a standard nonparametric rank test. In the course of this study, we have shown that the *TopRank* algorithm is a very efficient technique to detect several types of DoS attacks, as well as PortScan and NetScan. More precisely, the *TopRank* algorithm has two main features which make it very attractive. First, it is able to adapt to various types and intensities of traffic thanks to the nonparametric property of the test stage. Second, its computational simplicity and efficiency make its implementation feasible on the fly. This on-the-fly implementation has been tested in the context of an experimental detection system that we have jointly developed with France-Télécom and the other partners involved in the ANR-RNRT OSCAR project.

**Acknowledgments.**   The authors would like to thank the ANR (French National Research Agency) for their funding.

46, RUE BARRAULT
75634 PARIS CEDEX 13
FRANCE
E-MAIL: celine.levy-leduc@telecom-paristech.fr
        roueff@telecom-paristech.fr